# Evolution of Dopant-Concentration-Induced Magnetic Exchange Interaction in Topological Insulator Thin Films

Fei Wang[1, 2], Yi-Fan Zhao[1], Zijie Yan[1], Hemian Yi[1], Wei Yuan[1], Lingjie Zhou[1], Weiwei Zhao[2], Moses H. W. Chan[1], and Cui-Zu Chang[1]

[1]Department of Physics, The Pennsylvania State University, University Park, PA16802, USA

[2]School of Material Science and Engineering, Harbin Institute of Technology, Shenzhen 518055, China

Corresponding authors: cxc955@psu.edu (C.-Z. C.).

**Abstract: Two essential ingredients for the quantum anomalous Hall (QAH) effect, i.e. topological and magnetic orders, can be combined by doping magnetic ions into a topological insulator (TI) film. Through this approach, the QAH effect has been realized in chromium (Cr)- and/or vanadium (V)-doped TI $(Bi,Sb)_2Te_3$ thin films. In this work, we synthesize both V- and Cr-doped $Bi_2Te_3$ thin films with controlled dopant concentration using molecular beam epitaxy (MBE). By performing magneto-transport measurements, we find that both systems show an unusual but yet similar ferromagnetic response with respect to magnetic dopant concentration, specifically the Curie temperature does not increase monotonically but shows a local maximum at a critical dopant concentration. Our angle-resolved photoemission spectroscopy (ARPES) measurements show that the Cr/V doping introduces hole carriers into $Bi_2Te_3$, which consequently move the chemical potential toward the charge neutral point. In addition, the Cr/V doping also reduces the spin-orbit coupling of $Bi_2Te_3$ which drives it from a nontrivial TI to a trivial semiconductor. The unusual ferromagnetic response observed in Cr/V-doped $Bi_2Te_3$ thin films is attributed to the dopant-concentration-**

**induced magnetic exchange interaction, which displays the evolution from the van Vleck-type ferromagnetism in a nontrivial magnetic TI to the Ruderman-Kittel-Kasuya-Yosida (RKKY)-type ferromagnetism in a trivial diluted magnetic semiconductor. Our work provides insights into the ferromagnetic properties of magnetically doped TI thin films and facilitates the pursuit of high-temperature QAH effect.**

**Main text:** The quantum anomalous Hall (QAH) effect is a topological quantum state with quantized Hall resistance and zero longitudinal resistance in the absence of external magnetic fields [1-4]. The dissipationless chiral edge states of the QAH insulators provide us with a unique opportunity to develop next-generation electronics, spintronics, and topological quantum computations with low-energy consumption [5-7]. To realize a QAH insulator, the material system must possess both band topology and magnetism [8]. The most convenient approach for achieving these two essential ingredients is by doping magnetic ions into a topological insulator (TI) [9,10]. In magnetically doped TIs, the exchange interaction mechanism is analogous to that in conventional diluted magnetic semiconductors [11,12]. Through this approach, the QAH effect has been realized to date in two magnetically doped TI systems, i.e. Cr- and/or V-doped $(Bi,Sb)_2Te_3$ [9,10,13-15]. In these two systems, Cr and V atoms are magnetic impurities that substitute the Bi/Sb sites [16] (Fig. 1a). The formation of the ferromagnetic order breaks time-reversal symmetry and opens a magnetic exchange gap at the Dirac point of TI. When the Fermi level is tuned into the magnetic exchange gap, the magnetic TI becomes a QAH insulator. To date, the critical temperature of the QAH state observed in these magnetically doped TI films/heterostructures is still limited to ~10 K [13,17]. Studies of the ferromagnetic response in Cr- and/or V-doped $Bi_2Te_3$ as a function of the dopant concentration can help us understand the magnetic exchange interactions and thus opens the exploration of the higher temperature QAH insulators.

Two common magnetic exchange interactions have been proposed to be responsible for the formation of the ferromagnetic order in magnetically doped TIs. One is the van Vleck-type exchange interaction, which originates from the nonzero matrix elements of the spin operators between the bulk conduction and valence bands [4,18]. The local magnetic moments are coupled without the assistance of itinerant carriers (Fig. 1b). The van Vleck-type exchange interaction relies on the strength of spin-orbit coupling (SOC) and is usually significantly enhanced by the appearance of an inverted band structure in a material. The van Vleck-type exchange interaction is thought to be a prerequisite for the realization of the QAH effect in magnetically doped TIs [4,9,10,14,15]. In addition, since the magnetically doped TI is also a special kind of diluted magnetic semiconductor, its ferromagnetic response can also be mediated through itinerant carriers by the Ruderman-Kittel-Kasuya-Yosida (RKKY) exchange interaction [11,12] (Fig. 1c). In particular, when the magnetic dopant concentration is high, the SOC of magnetically doped TI will be greatly reduced. This will make the inverted band structure fade away and the RKKY-type ferromagnetism will take over in magnetically doped TIs [5]. In most conventional diluted magnetic semiconductors, the Curie temperature shows a monotonic increase with dopant concentration [11,12]. For a fixed dopant concentration, the ferromagnetic order can be altered by tuning the carrier density and/or electric field [19-21].

In this work, we use molecular beam epitaxy (MBE) to synthesize V- and Cr-doped $Bi_2Te_3$ films with different dopant concentrations and perform *in-situ* angle-resolved photoemission spectroscopy (ARPES) and *ex-situ* electrical transport measurements on these films. ARPES results show that the V/Cr doping introduces hole carriers in $Bi_2Te_3$ films and moves their chemical potential towards the charge neutral point. Transport results show that the Curie temperature of V/Cr-doped $Bi_2Te_3$ films does not increase monotonically and shows a local maximum at a critical dopant concentration. Since the Cr/V doping also reduces the SOC of $Bi_2Te_3$, heavy Cr/V doping can drive it from a nontrivial TI to a trivial semiconductor [22-25]. We attribute the unusual ferromagnetic response observed in Cr/V-doped $Bi_2Te_3$ thin films as an evolution of the dopant-concentration-induced magnetic exchange interaction, i.e. from the van Vleck-type ferromagnetism in a nontrivial TI that crosses over to the RKKY-type ferromagnetism in a trivial diluted magnetic semiconductor.

All $(Bi_{1-x}V_x)_2Te_3$ ($0 \leq x \leq 0.23$) and $(Bi_{1-y}Cr_y)_2Te_3$ ($0 \leqslant y \leqslant 0.29$) films used in this work are synthesized in an ultra-high vacuum MBE chamber (Omicron Lab 10), which is connected to an ARPES chamber. Both chambers have a base vacuum ~$2 \times 10^{-10}$ mbar. The *in-situ* ARPES measurements are carried out using a Scientia R3000 analyzer with an unpolarized He-Iα light (~21.2 eV). Our MBE-grown V/Cr-doped $Bi_2Te_3$ films are characterized by atomic force microscopy (AFM), scanning tunneling microscopy (STM), X-ray diffraction (XRD), and ARPES measurements (Figs. S1 to S3) [26]. The electrical transport studies are carried out in a Physical Property Measurements System (Quantum Design, 2 K, 9 T) with the magnetic field applied perpendicular to the film plane. Six terminal mechanically-defined Hall bars are used for electrical transport measurements.

We first perform ARPES measurements on undoped $Bi_2Te_3$ (Fig. 1d), V-doped $Bi_2Te_3$(Fig. 1e), and Cr-doped $Bi_2Te_3$ (Figs. 1f and S3) films with a thickness of 10 quintuple layers (QLs)[26]. For undoped $Bi_2Te_3$, gapless Dirac surface states (SSs) are buried in the bulk valence bands. The Dirac point is located at ($E$-$E_F$) ~-250 meV, reflecting the electron-type carriers [27,28]. This observation is in agreement with the negative slope of the Hall traces (Figs. S5a, S8a, and S11a) [26]. At a small V/Cr concentration in $Bi_2Te_3$, the gapless Dirac SSs still appear but the chemical potential moves closer to the Dirac point. The Dirac point is located at $(E-E_F)$ ~ -148 meV for the $x$=0.05 sample (V doping) (Fig. 1e) and ~ -142 meV for the $y$=0.07 sample (Cr doping) (Fig. 1f). Therefore, the V/Cr ions act as hole donors in these V/Cr-doped $Bi_2Te_3$ films. With further increase of the V/Cr dopant concentrations, a large number of hole carriers are introduced, which move the chemical potential across the bulk valence bands. In addition, the Dirac point is pulled out of the bulk valence bands with V/Cr doping (Figs. 1e, 1f, and S3) [26].

Next, we carry out electrical transport measurements on both 10 QL $(Bi_{1-x}V_x)_2Te_3$ ($0 \leq x \leq 0.23$) and 6 QL $(Bi_{1-y}Cr_y)_2Te_3$ ($0 \leq y \leq 0.29$) films (Figs. 2 and 3). Figures 2a to 2g show the temperature dependence of the longitudinal resistance $\rho_{xx}$ of 10 QL $(Bi_{1-x}V_x)_2Te_3$ films with different V dopant concentrations $x$. For the $x$=0.02 sample, the $\rho_{xx}$ value decreases with decreasing $T$, showing a metallic behavior. At $T$ ~8.8 K, a hump feature is observed (Fig. 2a). Such behavior in the $\rho_{xx}$-$T$ curve of magnetic materials has been associated with the paramagnetism-to-ferromagnetism phase transition, primarily due to spin-disorder scattering [29,30]. This feature is often used to determine the magnetic ordering temperature (i.e. Curie temperature $T_C$) of the ferromagnetic materials [29,30]. We next gradually increase $x$ in V-doped $Bi_2Te_3$ films. For the $x \leq 0.09$ samples, the hump feature in $\rho_{xx}$-$T$ curves moves towards higher temperatures. The values of $T_C$ determined by the hump feature are ~10.2 K for $x$=0.05, ~15.1 K for $x$=0.07, and ~15.9 K for $x$=0.09 samples (Figs. 2b to 2d). In addition, we also find that the samples tend to show an insulating behavior with increasing $x$ in this regime, which is likely a result of the neutralization between the hole carriers introduced by the V doping and the original electron carriers in intrinsic $Bi_2Te_3$ (Fig. 1e). With a further increase in $x$, the samples still show an insulating behavior but the $\rho_{xx}$ value at $T$ =2 K decreases. For the $x$=0.13 sample, $T_C$ is around 10.4 K (Fig. 2e). The lower $T_C$ with a larger $x$ is unusual, and unlike the behavior in the most conventional diluted magnetic semiconductors [11,12]. For the $x>0.13$ samples, the $T_C$ value increases with increasing $x$ again. $T_C$ is found to be ~16.8 K for $x=0.18$ and ~20.0 K for $x=0.23$ samples (Figs. 2f and 2g).

To validate this unusual $T_C$ response in V-doped $Bi_2Te_3$ films, we perform anomalous Hall (AH) measurements on these 10 QL $(Bi_{1-x}V_x)_2Te_3$ ($0 \leq x \leq 0.23$) films under different temperatures. The $T_C$ value can be determined by the temperature at which the AH resistance at zero magnetic field $\rho_{yx}(0)$ vanishes. At $T=2$ K, all V-doped $Bi_2Te_3$ samples show square-shaped hysteresis loops, demonstrating the long-range ferromagnetic order with perpendicular anisotropy (Fig. S5)[26]. These ferromagnetic hysteresis loops shrink and show a linear $\rho_{yx}$-$\mu_0H$ dependence at ~$T_C$ of the samples. We plot $\rho_{yx}(0)$ as a function of $T$ and find that the $T_C$ value determined by AH measurement is slightly higher than that determined by the hump feature in its corresponding $\rho_{xx}$-$T$ curve (Figs. 2h to 2n). This difference might be a result of insufficient temperature points at which we perform the AH measurements and fail to spot the exact temperature where $\rho_{yx}(0)$ becomes zero. Nevertheless, the $T_C$ values determined by the $\rho_{yx}(0)$-$T$ curves also show an increase-reduction-increase behavior with a local maximum for the $x=0.09$ sample (Figs. 2h to 2n).

A similarly unusual ferromagnetic response is also observed in Cr-doped $Bi_2Te_3$ films. Compared to 10 QL $(Bi_{1-x}V_x)_2Te_3$ ($0 \leq x \leq 0.23$) films, the diminished bulk conduction contribution in 6 QL $(Bi_{1-y}Cr_y)_2Te_3$ ($0 \leqslant y \leqslant 0.29$) films smear the hump feature in their $\rho_{xx}$-$T$ curves (Fig. S7) [26]. Therefore, we perform AH measurements on 6 QL $(Bi_{1-y}Cr_y)_2Te_3$ ($0 \leqslant y \leqslant 0.29$) films with a narrow temperature interval ~1 K near $T_C$ (Figs. S8) [26]. Figures 3a to 3g show the temperature-dependent $\rho_{yx}(0)$ of 6 QL $(Bi_{1-y}Cr_y)_2Te_3$ films with different $y$. $T_C$ determined by the temperature at which $\rho_{yx}(0)$ becomes 0 is ~9 K for $y=0.05$, ~15 K for $y=0.07$, ~22 K for $y=0.10$, ~30 K for $y=0.12$,

~26 K for $y$=0.14, ~19 K for $y$=0.17, and ~27 K for $y$=0.22 samples. The $T_C$ values of Cr-doped Bi$_2$Te$_3$ films also show a clear increase-reduction-increase behavior with a local maximum for the $y$=0.12 sample. We note that the unusual ferromagnetic response is also observed in one more batch of 5 QL Cr-doped Bi$_2$Te$_3$ films (Figs. S10 to S14) [26].

To summarize the unusual ferromagnetic response in V/Cr-doped Bi$_2$Te$_3$ films, we plot their $T_C$ values and $\rho_{xx}$ values at $T$ =2 K as a function of magnetic dopant concentration $x(y)$ (Figs. 4a to 4d). As shown above, the $T_C$ value of the V-doped Bi$_2$Te$_3$ films increases with $x$ for $0 \leq x \leq 0.09$, decreases with $x$ for $0.09 \leq x \leq 0.13$, and increases again with $x$ for $x \geq 0.13$ (Fig. 4a). We find that $\rho_{xx}$ shows a peak at $x = 0.13$ (Fig. 4b), concomitant with a local minimum of $T_C$. The Cr-doped Bi$_2$Te$_3$ films show similar behaviors. $T_C$ increases with $y$ for $0 \leq y \leq 0.12$, decreases with $y$ for $0.12 \leq y \leq 0.17$, and increases again with $y$ for $y \geq 0.17$ (Fig. 4c). Similar to the behavior of the V-doped films, $\rho_{xx}$ shows a peak where a local minimum of $T_C$ is found at $y$ ~0.17 (Fig. 4d). As noted above, the V/Cr doping inevitably introduces hole carriers into Bi$_2$Te$_3$, which consequently move the chemical potential toward the charge neutral point. Heavy V/Cr doping may make the chemical potential cross the bulk valence bands. This explains why $\rho_{xx}$ shows a peak with varying $x$ or $y$. We note that the maximum of $\rho_{xx}$ usually corresponds to the charge neutral point [28,31].

Finally, we propose a phenomeonological model to understand the physical origin of the unusual ferromagnetic response in V- and Cr-doped Bi$_2$Te$_3$ systems. In addition to the introduction of hole carriers, as discussed above, the V/Cr doping also reduces the SOC of Bi$_2$Te$_3$ and thus drives it from a nontrivial TI to a trivial semiconductor [22-24]. For low V/Cr doping, i.e. $0 \leq x \leq 0.09$ for V and $0 \leq y \leq 0.12$ for Cr, V/Cr-doped Bi$_2$Te$_3$ film is still located in the nontrivial TI regime (Fig. 4e). Its inverted band structure, particularly when its chemical potential crosses Dirac SSs, generates considerable spin susceptibility and thus the van Vleck-type exchange interaction is dominant in V/Cr-doped $Bi_2Te_3$ films. Therefore, the Curie temperature $T_C$ increases because the V/Cr ions in $Bi_2Te_3$ introduce more electronic states in the total spin susceptibility [4,18], while $\rho_{xx}$ increases due to the reduced electron carrier density. We note that since the maximum of the bulk valence bands along Γ-M direction is usually higher than the Dirac point in undoped $Bi_2Te_3$ [27,28,32,33], a minor hole carrier-mediated RKKY exchange interaction may appear when the chemical potential starts to cross the bulk valence bands [34,35]. For high V/Cr doping, i.e. $x \geq 0.13$ for V doping and $y \geq 0.17$ for Cr doping, V/Cr-doped $Bi_2Te_3$ film becomes a trivial semiconductor and the inverted band structure disappears (Fig. 4e). The V/Cr-doped $Bi_2Te_3$ film is then simply a conventional diluted magnetic semiconductor, in which the van Vleck-type exchange interaction is negligible and the RKKY-type exchange interaction is dominant.

For the $Bi_2Te_3$ film with intermediate V/Cr doping, i.e. $0.09 \leq x \leq 0.13$ for V doping and $0.12 \leq y \leq 0.17$ for Cr doping, the van Vleck- and RKKY-type exchange interactions coexist and compete with each other [34-36]. In this region, an increase of V/Cr doping greatly reduces the carrier density and thus weakens the contribution of the RKKY-type exchange interaction. Moreover, the increase of V/Cr doping further reduces the SOC of $Bi_2Te_3$, which drives it to behave as a trivial semiconductor [26], so the van Vleck-type exchange interaction tends to be less favored. Therefore, the total magnetic exchange interaction becomes weak and thus $T_C$ decreases with increasing magnetic dopant concentration $x$ or $y$. Based on our above analysis, the V/Cr dopant-concentration-induced quantum transition between nontrivial TI and trivial semiconductor phases likely occurs near $x = 0.09$ for V doping and $y = 0.12$ for Cr doping. The latter value for the Cr-doped $Bi_2Te_3$ system agrees well with that from our recent studies on QAH multilayer heterostructures [24,25].

To summarize, we fabricate a series of V/Cr-doped $Bi_2Te_3$ films with different V/Cr dopant concentrations and find an unusual ferromagnetic response in these films, i.e. a local maximum of the Curie temperature $T_C$ at critical V/Cr dopant concentrations. By careful analysis, we attribute this unusual ferromagnetic response in V/Cr-doped $Bi_2Te_3$ systems to an evolution of the magnetic exchange interaction, i.e. the crossover from van Vleck-type exchange interaction in the nontrivial TI regime to RKKY-type interaction in the trivial semiconductor regime. We conclude that the unusual ferromagnetic response of the V/Cr-doped $Bi_2Te_3$ films is due primarily to the two roles of the V/Cr doping: (*i*) it introduces a large number of hole carriers and moves the chemical potential pass through the charge neutral point and cross the bulk valence bands; (*ii*) it reduces the SOC of $Bi_2Te_3$ and drives it from a nontrivial TI to a trivial semiconductor. Our work provides insights into the magnetic exchange interactions in magnetically doped TI films/heterostructures and lays down the requirements and constraints on how to pursue the higher temperature QAH effect and other novel topological phases therein.

**Acknowledgments:** We thank Chao-Xing Liu, Nitin Samarth, Xuepeng Wang, Di Xiao, and Haijun Zhang for their helpful discussions. This work is primarily supported by the NSF-CAREER award (DMR-1847811), including the MBE growth and electrical transport measurements. The ARPES measurements and sample characterization are partially supported by the AFOSR grant (FA9550-21-1-0177). C. Z. C. also acknowledges the support from the Gordon and Betty Moore Foundation's EPiQS Initiative (Grant GBMF9063 to C. -Z. C.). W.W. Z. acknowledges the support from the Natural Science Foundation of China (52073075) and Shenzhen Science and Technology Program (KQTD20170809110344233).

**Figures and figure captions:**

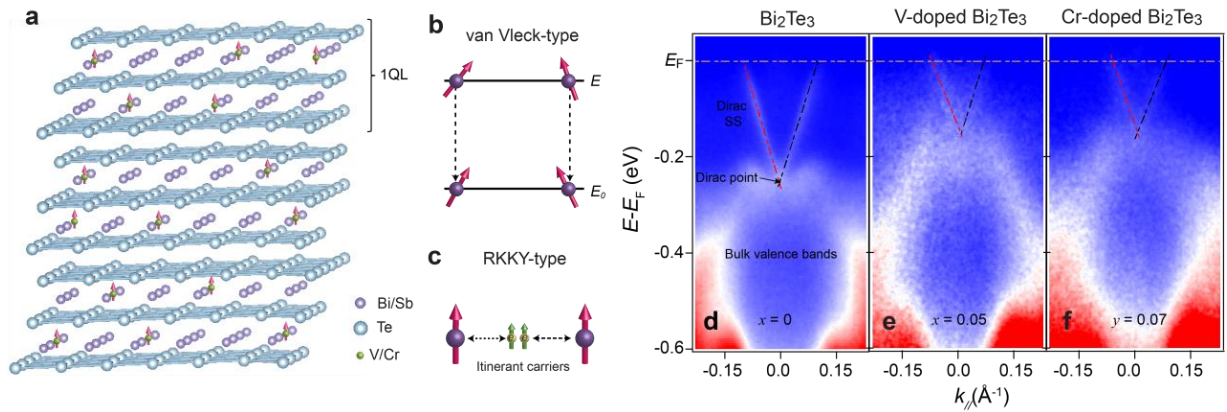

**Fig. 1| Two types of exchange interaction in magnetically doped TI films.** (a) Schematic atomic structure of the Cr- or V-doped (Bi,Sb)$_2$Te$_3$ with the QAH state. Cr/V substitutes for the Bi/Sb sites. (b) van Vleck-type magnetic exchange interaction. The van Vleck-type ferromagnetism is induced by the mixing of the bulk conduction and valences bands and thus can be significantly enhanced in a system with the inverted band structure. It is usually insensitive to itinerant carriers. (c) RKKY-type magnetic exchange interaction. In RKKY-type ferromagnetism, local magnetic moments are coupled through itinerant carriers. (d-f) ARPES band maps of 10 QL Bi$_2$Te$_3$ (d), (Bi$_{1-x}$V$_x$)$_2$Te$_3$ ($x$=0.05) (e), and (Bi$_{1-y}$Cr$_y$)$_2$Te$_3$ ($y$=0.07) (f). All these samples are grown on the bilayer graphene-terminated 6H-SiC(0001) substrates. All ARPES measurements are performed at room temperature. The chemical potential of Bi$_2$Te$_3$ films moves towards the Dirac point, indicating hole-type carriers are introduced by doping Cr/V.

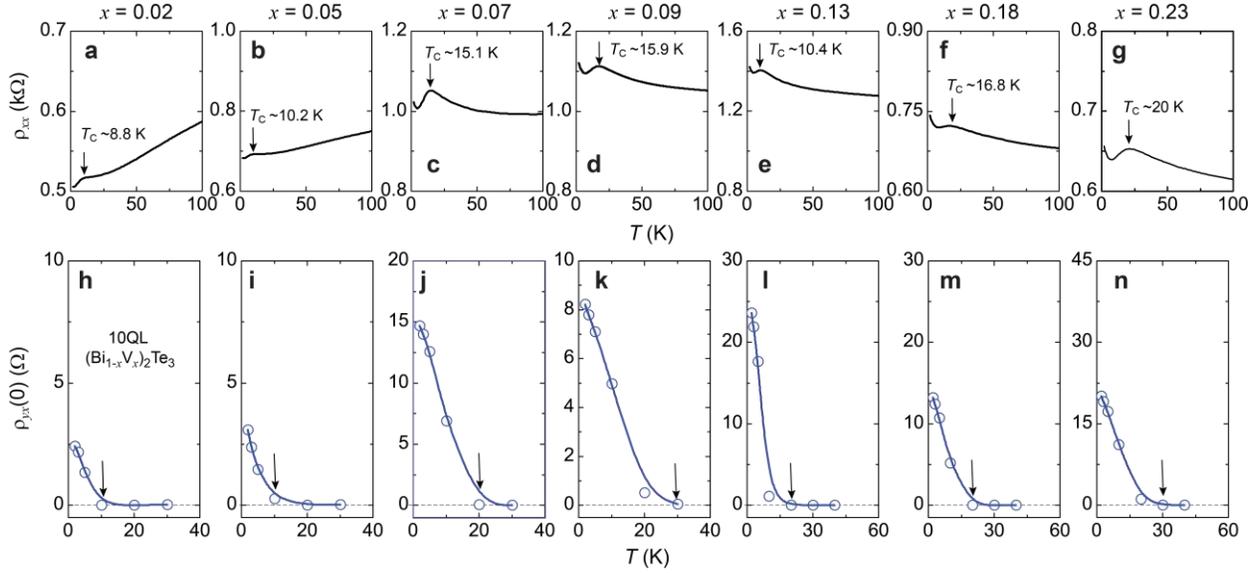

**Fig. 2| Transport results of 10 QL $(Bi_{1-x}V_x)_2Te_3$ films with different V dopant concentration $x$.** (a-g) Temperature dependence of the longitudinal resistance $\rho_{xx}$ of 10 QL $(Bi_{1-x}V_x)_2Te_3$ films. (a) $x=0.02$; (b) $x=0.05$; (c) $x=0.07$; (d) $x=0.09$; (e) $x=0.13$; (f) $x=0.18$; (g) $x=0.23$. The hump features marked by arrows indicate the Curie temperature $T_C$. (h-n) Temperature dependence of the zero magnetic field Hall resistance $\rho_{yx}(0)$ of 10 QL $(Bi_{1-x}V_x)_2Te_3$ films. (h) $x=0.02$; (i) $x=0.05$; (j) $x=0.07$; (k) $x=0.09$; (l) $x=0.13$; (m) $x=0.18$; (n) $x=0.23$. The arrows indicate the critical temperatures at which $\rho_{yx}(0)$ is vanishing.

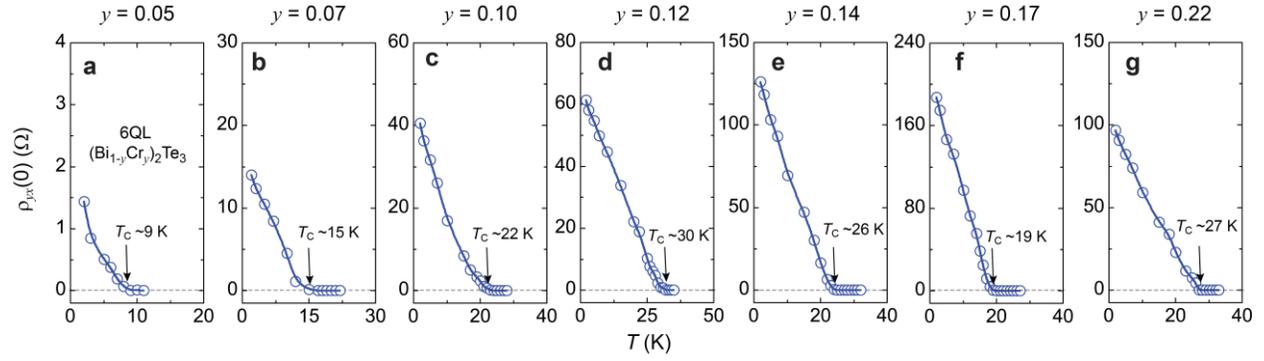

**Fig. 3| Transport results of 6 QL $(Bi_{1-y}Cr_y)_2Te_3$ films with different Cr dopant concentration $y$.** (a-g) Temperature dependence of $\rho_{yx}(0)$ of 6 QL $(Bi_{1-y}Cr_y)_2Te_3$ films. (a) $y=0.05$; (b) $y=0.07$; (c) $y=0.10$; (d) $y=0.12$; (e) $y=0.14$; (f) $y=0.17$; (g) $y=0.22$. The arrows indicate the critical temperatures at which $\rho_{yx}(0)$ vanishes.

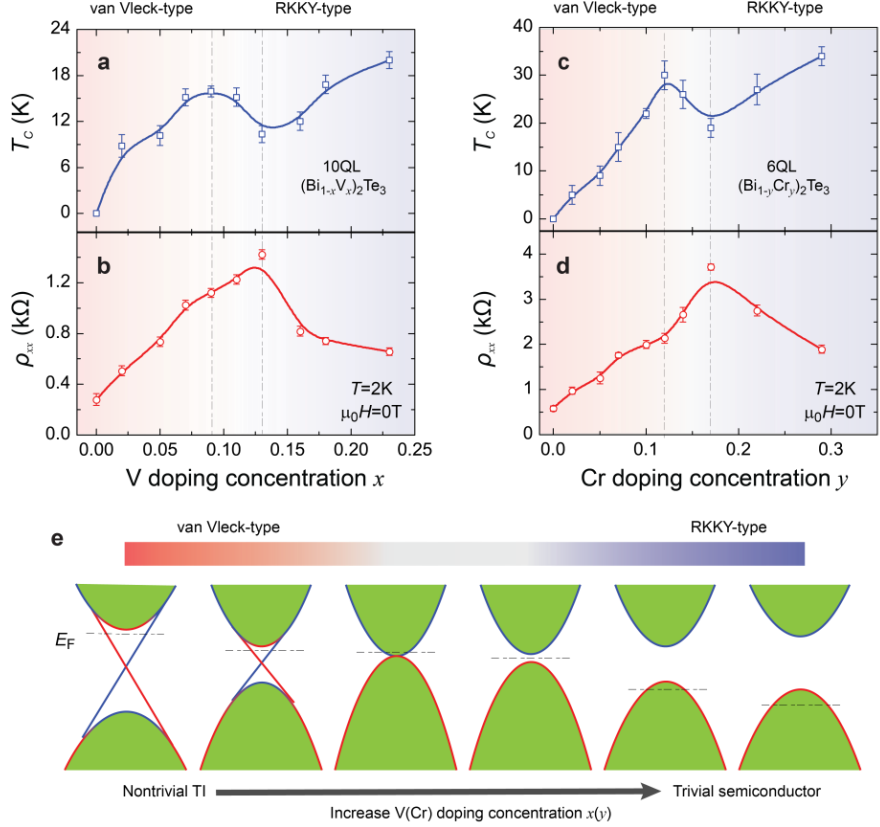

**Fig. 4| Evolution of dopant-concentration-induced exchange interaction in magnetically doped Bi$_2$Te$_3$.** (a,b) V dopant concentration $x$-dependent $T_C$ (a) and $\rho_{xx}$ (b) of 10 QL (Bi$_{1-x}$V$_x$)$_2$Te$_3$ films. (c,d) Cr dopant concentration $y$-dependent $T_C$ (a) and $\rho_{xx}$ (b) of 6 QL (Bi$_{1-y}$Cr$_y$)$_2$Te$_3$ films. The $\rho_{xx}$ values are acquired at $T = 2$ K. (e) Schematics of the evolution of the electronic band structure (i.e. nontrivial to trivial) and the chemical potential position (i.e. electron- to hole-doped) with increasing V(Cr) dopant concentration $x(y)$.

**References:**

[1]     F. D. M. Haldane, Model for a Quantum Hall-Effect without Landau Levels: Condensed-Matter Realization of the "Parity Anomaly", *Phys. Rev. Lett.* **61**, 2015 (1988).

[2]     C. X. Liu, X. L. Qi, X. Dai, Z. Fang, and S. C. Zhang, Quantum anomalous Hall effect in $Hg_{1-y}Mn_yTe$ quantum wells, *Phys. Rev. Lett.* **101**, 146802 (2008).

[3]     X. L. Qi, T. L. Hughes, and S. C. Zhang, Topological Field Theory of Time-Reversal Invariant Insulators, *Phys. Rev. B* **78**, 195424 (2008).

[4]     R. Yu, W. Zhang, H. J. Zhang, S. C. Zhang, X. Dai, and Z. Fang, Quantized Anomalous Hall Effect in Magnetic Topological Insulators, *Science* **329**, 61 (2010).

[5]     C. Z. Chang and M. D. Li, Quantum Anomalous Hall Effect in Time-Reversal-Symmetry Breaking Topological Insulators, *J. Phys. Condens. Mat.* **28**, 123002 (2016).

[6]     H. M. Weng, R. Yu, X. Hu, X. Dai, and Z. Fang, Quantum Anomalous Hall Effect and Related Topological Electronic States, *Adv. Phys.* **64**, 227 (2015).

[7]     Y. Tokura, K. Yasuda, and A. Tsukazaki, Magnetic topological insulators, *Nat. Rev. Phys.* **1**, 126 (2019).

[8]     C. Z. Chang, Marriage of topology and magnetism, *Nat. Mater.* **19**, 484 (2020).

[9]     C. Z. Chang, J. S. Zhang, X. Feng, J. Shen, Z. C. Zhang, M. H. Guo, K. Li, Y. B. Ou, P. Wei, L. L. Wang, Z. Q. Ji, Y. Feng, S. H. Ji, X. Chen, J. F. Jia, X. Dai, Z. Fang, S. C. Zhang, K. He, Y. Y. Wang, L. Lu, X. C. Ma, and Q. K. Xue, Experimental Observation of the Quantum Anomalous Hall Effect in a Magnetic Topological Insulator, *Science* **340**, 167 (2013).

[10]    C. Z. Chang, W. W. Zhao, D. Y. Kim, H. J. Zhang, B. A. Assaf, D. Heiman, S. C. Zhang, C. X. Liu, M. H. W. Chan, and J. S. Moodera, High-Precision Realization of Robust Quantum Anomalous Hall State in a Hard Ferromagnetic Topological Insulator, *Nat. Mater.* **14**, 473 (2015).

[11]    T. Jungwirth, J. Sinova, J. Masek, J. Kucera, and A. H. MacDonald, Theory of ferromagnetic (III,Mn)V semiconductors, *Rev. Mod. Phys.* **78**, 809 (2006).

[12]    T. Dietl and H. Ohno, Dilute ferromagnetic semiconductors: Physics and spintronic structures, *Rev. Mod. Phys.* **86**, 187 (2014).

[13]    Y. Ou, C. Liu, G. Jiang, Y. Feng, D. Zhao, W. Wu, X. X. Wang, W. Li, C. Song, L. L. Wang, W. Wang, W. Wu, Y. Wang, K. He, X. C. Ma, and Q. K. Xue, Enhancing the Quantum

Anomalous Hall Effect by Magnetic Codoping in a Topological Insulator, *Adv. Mater.* **30**, 1703062 (2017).

[14] J. G. Checkelsky, R. Yoshimi, A. Tsukazaki, K. S. Takahashi, Y. Kozuka, J. Falson, M. Kawasaki, and Y. Tokura, Trajectory of the Anomalous Hall Effect towards the Quantized State in a Ferromagnetic Topological Insulator, *Nat. Phys.* **10**, 731 (2014).

[15] X. F. Kou, S. T. Guo, Y. B. Fan, L. Pan, M. R. Lang, Y. Jiang, Q. M. Shao, T. X. Nie, K. Murata, J. S. Tang, Y. Wang, L. He, T. K. Lee, W. L. Lee, and K. L. Wang, Scale-Invariant Quantum Anomalous Hall Effect in Magnetic Topological Insulators beyond the Two-Dimensional Limit, *Phys. Rev. Lett.* **113**, 137201 (2014).

[16] C. Z. Chang, J. S. Zhang, M. H. Liu, Z. C. Zhang, X. Feng, K. Li, L. L. Wang, X. Chen, X. Dai, Z. Fang, X. L. Qi, S. C. Zhang, Y. Y. Wang, K. He, X. C. Ma, and Q. K. Xue, Thin Films of Magnetically Doped Topological Insulator with Carrier-Independent Long-Range Ferromagnetic Order, *Adv. Mater.* **25**, 1065 (2013).

[17] M. Mogi, R. Yoshimi, A. Tsukazaki, K. Yasuda, Y. Kozuka, K. S. Takahashi, M. Kawasaki, and Y. Tokura, Magnetic Modulation Doping in Topological Insulators toward Higher-Temperature Quantum Anomalous Hall Effect, *Appl. Phys. Lett.* **107**, 182401 (2015).

[18] M. D. Li, C. Z. Chang, L. J. Wu, J. Tao, W. W. Zhao, M. H. W. Chan, J. S. Moodera, J. Li, and Y. M. Zhu, Experimental Verification of the Van Vleck Nature of Long-Range Ferromagnetic Order in the Vanadium-Doped Three-Dimensional Topological Insulator $Sb_2Te_3$, *Phys. Rev. Lett.* **114**, 146802 (2015).

[19] H. Ohno, D. Chiba, F. Matsukura, T. Omiya, E. Abe, T. Dietl, Y. Ohno, and K. Ohtani, Electric-Field Control of Ferromagnetism, *Nature* **408**, 944 (2000).

[20] D. Chiba, F. Matsukura, and H. Ohno, Electric-field control of ferromagnetism in (Ga,Mn)As, *Appl. Phys. Lett.* **89**, 162505 (2006).

[21] D. Chiba, A. Werpachowska, M. Endo, Y. Nishitani, F. Matsukura, T. Dietl, and H. Ohno, Anomalous Hall Effect in Field-Effect Structures of (Ga,Mn)As, *Phys. Rev. Lett.* **104** (2010).

[22] C. Z. Chang, P. Z. Tang, Y. L. Wang, X. Feng, K. Li, Z. C. Zhang, Y. Y. Wang, L. L. Wang, X. Chen, C. X. Liu, W. H. Duan, K. He, X. C. Ma, and Q. K. Xue, Chemical-Potential-Dependent Gap Opening at the Dirac Surface States of $Bi_2Se_3$ Induced by Aggregated Substitutional Cr Atoms, *Phys. Rev. Lett.* **112**, 056801 (2014).15

[23] J. S. Zhang, C. Z. Chang, P. Z. Tang, Z. C. Zhang, X. Feng, K. Li, L. L. Wang, X. Chen, C. X. Liu, W. H. Duan, K. He, Q. K. Xue, X. C. Ma, and Y. Y. Wang, Topology-Driven Magnetic Quantum Phase Transition in Topological Insulators, *Science* **339**, 1582 (2013).

[24] Y. F. Zhao, R. Zhang, R. Mei, L. J. Zhou, H. Yi, Y. Q. Zhang, J. Yu, R. Xiao, K. Wang, N. Samarth, M. H. W. Chan, C. X. Liu, and C. Z. Chang, Tuning the Chern number in quantum anomalous Hall insulators, *Nature* **588**, 419 (2020).

[25] Y.-F. Zhao, R. Zhang, L.-J. Zhou, R. Mei, Z.-J. Yan, M. H. W. Chan, C.-X. Liu, and C.-Z. Chang, Zero Magnetic Field Plateau Phase Transition in Higher Chern Number Quantum Anomalous Hall Insulators, *arXiv:2109.11382* (2021).

[26] See Supplemental Material at XXXXX for further details regarding the MBE growth, sample characterizations, and more transport results.

[27] Y. L. Chen, J. G. Analytis, J. H. Chu, Z. K. Liu, S. K. Mo, X. L. Qi, H. J. Zhang, D. H. Lu, X. Dai, Z. Fang, S. C. Zhang, I. R. Fisher, Z. Hussain, and Z. X. Shen, Experimental Realization of a Three-Dimensional Topological Insulator, $Bi_2Te_3$, *Science* **325**, 178 (2009).

[28] J. S. Zhang, C. Z. Chang, Z. C. Zhang, J. Wen, X. Feng, K. Li, M. H. Liu, K. He, L. L. Wang, X. Chen, Q. K. Xue, X. C. Ma, and Y. Y. Wang, Band Structure Engineering in $(Bi_{1-x}Sb_x)_2Te_3$ Ternary Topological Insulators, *Nat. Commun.* **2**, 574 (2011).

[29] Z. H. Zhou, Y. J. Chien, and C. Uher, Thin film dilute ferromagnetic semiconductors $Sb_{2-x}Cr_xTe_3$ with a Curie temperature up to 190 K, *Phys. Rev. B* **74**, 224418 (2006).

[30] C. Z. Chang, M. H. Liu, Z. C. Zhang, Y. Y. Wang, K. He, and Q. K. Xue, Field-effect modulation of anomalous Hall effect in diluted ferromagnetic topological insulator epitaxial films, *Sci. China Phys. Mech.* **59**, 637501 (2016).

[31] C. Z. Chang, Z. C. Zhang, K. Li, X. Feng, J. S. Zhang, M. H. Guo, Y. Feng, J. Wang, L. L. Wang, X. C. Ma, X. Chen, Y. Y. Wang, K. He, and Q. K. Xue, Simultaneous Electrical-Field-Effect Modulation of Both Top and Bottom Dirac Surface States of Epitaxial Thin Films of Three-Dimensional Topological Insulators, *Nano Lett.* **15**, 1090 (2015).

[32] Y. Y. Li, G. A. Wang, X. G. Zhu, M. H. Liu, C. Ye, X. Chen, Y. Y. Wang, K. He, L. L. Wang, X. C. Ma, H. J. Zhang, X. Dai, Z. Fang, X. C. Xie, Y. Liu, X. L. Qi, J. F. Jia, S. C. Zhang, and Q. K. Xue, Intrinsic Topological Insulator $Bi_2Te_3$ Thin Films on Si and Their Thickness Limit, *Adv. Mater.* **22**, 4002 (2010).

[33]   D. S. Kong, Y. L. Chen, J. J. Cha, Q. F. Zhang, J. G. Analytis, K. J. Lai, Z. K. Liu, S. S. Hong, K. J. Koski, S. K. Mo, Z. Hussain, I. R. Fisher, Z. X. Shen, and Y. Cui, Ambipolar field effect in the ternary topological insulator $(Bi_xSb_{1-x})_2Te_3$ by composition tuning, *Nat. Nanotechnol.* **6**, 705 (2011).

[34]   X. F. Kou, M. R. Lang, Y. B. Fan, Y. Jiang, T. X. Nie, J. M. Zhang, W. J. Jiang, Y. Wang, Y. G. Yao, L. He, and K. L. Wang, Interplay between Different Magnetisms in Cr-Doped Topological Insulators, *ACS Nano* **7**, 9205 (2013).

[35]   Z. C. Zhang, X. Feng, M. H. Guo, K. Li, J. S. Zhang, Y. B. Ou, Y. Feng, L. L. Wang, X. Chen, K. He, X. C. Ma, Q. K. Xue, and Y. Y. Wang, Electrically Tuned Magnetic Order and Magnetoresistance in a Topological Insulator, *Nat. Commun.* **5**, 4915 (2014).

[36]   J. Kim, S. H. Jhi, A. H. MacDonald, and R. Q. Wu, Ordering mechanism and quantum anomalous Hall effect of magnetically doped topological insulators, *Phys. Rev. B* **96**, 140410 (2017).